\documentstyle[prl,aps,twocolumn]{revtex}
\large  

\input epsf
\begin{document}
\draft \twocolumn[\hsize\textwidth
\columnwidth\hsize\csname
@twocolumnfalse\endcsname
\title{How to split the electron in one dimension}
\author{Eugene B. Kolomeisky $^{(1)}$, T. J. Newman $^{(1,2)}$, 
and Joseph P. Straley$^{(3)}$} 
\address{$^{(1)}$ Department of Physics, University of Virginia,
P. O. Box 400714, Charlottesville, Virginia 22904-4714\\
$^{(2)}$ Department of Biology, University of Virginia, Charlottesville,
Virginia 22903\\
$^{(3)}$ Department of Physics and Astronomy, University of Kentucky,
Lexington, Kentucky 40506-0055}
\maketitle
\begin{abstract}
Using the example of the Davydov soliton - a large acoustic polaron in one 
dimension - we demonstrate that the electron wave function can be fissioned  
in two or more long-lived, well-localized and spatially arbitrarily far 
separated fragments.  The phenomenon of wave function splitting is a 
result of the electron-medium interaction, and takes place under a variety of 
conditions provided the initial wave function of the electron is localized 
and has at least one node.  
\end{abstract}
\vspace{2mm} 
]

\narrowtext

The concept of the wave function and the invention of quantum mechanics are 
among the most fascinating and  powerful achievements  of physics. The 
purpose of this Letter is to show yet another intriguing manifestation of
quantum mechanics, namely that the wave function of an elementary 
particle can be fissioned into well-localized, spatially well-separated, and 
long-lived pieces.

The idea of fissioning has been put forward by Maris in his analysis of the 
physics of electron bubbles in superfluid helium \cite{Maris}.  It has been 
known for years that an electron injected 
into liquid helium can lower its energy by self-localizing in a $1s$ state 
inside a spherical cavity from which helium atoms are expelled.  Maris argues 
that if the electron is optically excited from the $1s$ state to the $1p$ 
state, the pressure exerted by the electron in the excited state will no 
longer support the spherical cavity -- the bubble walls will be set into 
motion, and the inertia of the liquid around the bubble will suffice to break 
the bubble into two smaller bubbles each carrying half of the original wave 
function.  Maris further argues that the daughter bubbles will act in such a
way as if they were fractions of the original particle, and that the existence
of ``fractional'' bubbles explains a substantial amount of otherwise 
unexplained experimental data.

Elser \cite{Elser} disagrees, arguing that the failure of the adiabatic
approximation used in \cite{Maris} makes it impossible to fission the wave 
function. 

Jackiw, Rebbi and Schrieffer \cite{jrs} do not contest the existence of 
a bubble carrying ``half of the electron's wave function'' but point out that
an
electron in a helium bubble is entangled rather than fractional: before 
a
measurement one cannot state in which bubble the electron resides; after 
the measurement a full electron will be found in one bubble and nothing in the other.

The dispute initiated by the proposal \cite{Maris} raises a fundamental 
question: Is it in principle possible to fission the wave function of an 
elementary particle in long-lived localized well-separated fragments?  

We found an affirmative answer to this question in a related physical 
context.  Part of the difficulty in making definite statements about 
electron bubbles in helium is that the treatments involve too many 
approximations \cite{Maris}; some of which are hardly 
justifiable \cite{Elser}. At the same time it seems difficult to improve 
the approximations without making the problem unmanageable.

Electrons form bubbles in helium as a consequence of a kind of  
polaronic effect\cite{Rashba}: they become 
self-localized as a result of the response of the 
medium.
Instead we will look at a quantum particle in the presence of a deformable 
classical medium, a one-dimensional polaron problem.  Among the attractive
features of this system is that 
self-localization takes place regardless of the magnitude 
of the coupling between the electron and the medium
\cite{Rashba,Davydov}.
When the coupling is small, 
the scale of the self-localized state can be much bigger than the 
atomic spacing (``large'' polaron limit), and thus a macroscopic approach
that ignores
the discreteness of the medium can be adopted. From a practical 
viewpoint, polaronic effects play an important role in the physics 
of quasi-one-dimensional conductors \cite{Heeger}.   Polarons (also called 
Davydov solitons) are known to exist in $\alpha$ - helical protein molecules 
of living organisms, and play a crucial role in the transport of vibrational
energy and charge \cite{Davydov}.  These systems represent experimental 
candidates where the effects described below can be tested.

The dynamics of the system is described by the set of Davydov equations 
\cite{Davydov}:
  
\begin{equation}
\label{SE}
i \hbar \partial _{t} \Psi  
= -{\hbar ^{2} \over 2m} \partial _{x}^{2}\Psi + \lambda \partial_{x}u\Psi \ , 
\end{equation}
\begin{equation}
\label{WE}
\rho\partial^{2}_{t}u = B\partial^{2}_{x}u + \lambda\partial_{x}|\Psi|^{2} \ ,
\end{equation}
where $m$ is the electron mass, $\Psi(x,t)$ is the electron wave function 
normalized to unity, $u(x,t)$ is the classical displacement field of 
the medium, $B$ is the bulk constant of the medium, $\rho$ is the linear mass 
density of the medium, and $\lambda$ is the constant of electron-medium 
interaction.  The first of the equations (\ref{SE}) is the 
Schr\"odinger equation 
for the electron moving in a self-induced potential $\lambda \partial_{x}u$ 
due to the strain in the medium, while the second
equation (\ref{WE}) is the inhomogeneous wave equation with an extra term 
accounting for 
the force exerted by the electron on the particles of the medium. A classical 
description of the medium degrees of freedom is justified as the medium is 
composed of a large number of particles which are much heavier than the 
electron.  

The ground state of the coupled electron-medium system is described by the 
soliton solution to (\ref{SE}) and (\ref{WE}) of the form \cite{Davydov}:  
$\Psi_{0} = \exp( - iE_{0}t/\hbar){\rm sech}(x/\xi)/(2\xi)^{1/2}$ and
$\partial_{x}u_{0} = - (\lambda/B)|\Psi_{0}|^{2}$, where $E_{0} = 
- \hbar^{2}/(2m\xi^{2})$, and $\xi = 2B\hbar^{2}/(m\lambda^{2})$. The physical
meaning of this solution is that the electron deforms the medium, which acts 
as a potential well localizing the electron in its lowest energy state 
with wave function $\Psi_{0}$ and energy $E_{0} < 0$.   
Including the deformation energy, the 
total
energy change due to the formation of the soliton is $E_{0}/3 $, 
which is still negative.  Thus the soliton forms
spontaneously.  The location of the soliton center is arbitrary, and the 
localization length $\xi$ will be  assumed to be 
much larger than the atomic scale.  The Davydov soliton is
remotely analogous to the 
$1s$ electron bubble in helium \cite{Maris,Elser} -- the potential well 
$\lambda\partial_{x}u_{0} = - (\lambda^{2}/B)|\Psi_{0}|^{2} = 
-(\hbar^{2}/m\xi^{2}){\rm sech}^{2}(x/\xi)$ 
created and sensed by the electron is 
analogous to the cavity of expelled helium atoms confining the 
electron motion.

Using the dimensionless variables for length $y = x/\xi$, time 
$\tau = |E_{0}|t/\hbar$, the wave function $\Phi = \xi^{1/2}\Psi$, and the 
potential felt by the electron $w = \lambda\partial_{x}u/|E_{0}|$, the system 
of equations (\ref{SE}) and (\ref{WE}) simplifies to:    
        
\begin{equation}
\label{newSE}
i \partial_{\tau} \Phi  = -\partial _{y}^{2}\Phi + w\Phi \ ,
\end{equation}
\begin{equation}
\label{newWE}
s\partial^{2}_{\tau}w = \partial^{2}_{y}[w + 4|\Phi|^{2}] \ .
\end{equation}
It is interesting that the parameter 
$s = \lambda^{4}\rho/(16B^{3}\hbar^{2})$ controlling the strength of 
electron-medium coupling does not depend on the electron mass.  The 
dimensionless description (\ref{newSE}) and (\ref{newWE}) will be used 
hereafter. 

We are now ready to describe how to split the wave function into fragments.  
In the limit that a fragment is infinitely far away from other features of 
the wave function it is necessarily of the ${\rm sech}$ form and the 
supporting medium well is of the ${\rm sech}^{2}$ form.  
For example, half of the Davydov soliton is 
described by $\Phi_{1/2} = \exp(i\tau/4){\rm sech}(y/2)/2\surd{2}$ 
(normalized to one half) 
and $w_{1/2} = -{\rm sech}^{2}(y/2)/2$.  It is straightforward to calculate 
that the total energy gain from having in the system two infinitely far 
separated halves of the Davydov soliton is four times smaller than that from 
having one full soliton, showing that the Davydov soliton is a stable
object which does not spontaneously break in half.  However the soliton can be 
split if extra energy is invested into the system.

To prepare the initial state which will split into fragments one can try,
following Maris \cite{Maris}, to
optically excite the electron into the first excited state of the 
self-induced potential \cite{note}. 
However this idea will not work in one dimension: the self-induced potential
$w_{0} = -4|\Phi_{0}|^{2} = - 2{\rm sech}^{2}y$ has only one bound 
state\cite{Rashba}, \cite{LL1}.  
Nonetheless, it is possible to excite the electron into an initial state that  
is well-localized, even though it is a superposition of the unbound states,
such that the wave function separates into distinct and well-localized 
components.  For example, for the potential $w_{0} = - 2{\rm sech}^{2}y$ the 
zero-energy state is on the verge of becoming discrete \cite{LL1}.  Then 
imposing arbitrarily weak even external confining potential will turn the
marginal state into an odd localized excited state.  Exciting the electron 
into this newly formed 
state and turning off the external potential, we  arrive at the initial 
condition resembling the $1p$ electron inside the spherical helium bubble
\cite{Maris}. 

More generally, if initially the wave function is odd and well-localized, and 
the strain field is even and well-localized, the zero of the wave function 
persists through the evolution.  This node plays the role of an impenetrable 
wall that repels the wings of the wave function.  Without the coupling with 
the medium there would be a long-range repulsion between the wings and the 
node and the wave function will just spread out away from the origin.  The 
presence of the medium prevents spreading by localizing the wave function and  
making its interaction with the node short ranged.  After the initial
breakup the fragments of the wave function will start aggregating into the 
appropriate  ${\rm sech}$ form, and will virtually stop feeling the presence of
the node as soon as they are several localization lengths apart from the 
origin.  At the same time the material degrees of freedom will keep adjusting 
into the appropriate ${\rm sech^{2}}$ form.  This process of "settling down" 
into the Davydov form will be accompanied by the fragments emitting sound 
waves which establishes a "communication" between the pieces even when
they are far apart.  The exchange of sound waves leads to a 
very long-range repulsion between the fragments, since the 
amplitude of sound waves is a constant independent of the distance to the
source of sound (in one dimension).
Thus the combination of inertia acquired as a result of
the
initial breakup, and the long-range repulsion through exchange by sound will 
separate the fragments arbitrarily far from each other.           

For the initial conditions described above the exchange of probabilities 
between the fragments is forbidden by symmetry.  In reality there can be 
deviations from the "ideal" conditions.  Then the initial zero of the
wave function will not be preserved by the evolution -- fragments of 
unequal sizes can start exchanging by a flow of probability density.  This 
might 
lead to the larger fragment absorbing the smaller one, thus restoring 
the full Davydov soliton in a microscopic time.  To understand this case we 
resort to numerical experiments which are described below.

We tested these ideas by numerically solving the system of equations 
(\ref{newSE}) and (\ref{newWE}) under a variety of initial conditions.  
We use a Fourier transform algorithm to integrate forward the Schr\"odinger
equation, thus maintaining unitary evolution perfectly. The wave equation for 
the stress field is integrated forward using the Verlet algorithm\cite{Verlet}.
Figures 1 and 2 show "carpets" of the electron density $|\Phi|^2$ 
and the well function $w$ as functions of position and time. The horizontal 
direction corresponds to space while the vertical to time; the 
time arrow runs from the bottom to the top, and the bottom of the Figures 
represents the initial conditions.  The electron
density is strictly non-negative and shown using varying shades of red -- the 
higher the density, the deeper the color.  The well function is mainly 
negative and shown using varying shades of blue -- the intensity of the color 
is proportional to the well depth.  The places where the well function becomes
positive are color-coded in red with the same intensity-magnitude 
correspondence.  
The Figures are constructed for $s = 1$ and 
 $\Phi (y,0) \sim \exp[-0.5(y - y_{c})^{2}/d^{2}]{\rm tanh}y$ -- 
for $y_{c} = 0$ this is an odd function localized near the origin
($d = 30$ was selected), while the parameter $y_{c}$ controls the
asymmetry
of the initial electron density.
The Figures are produced for $w(y,0) = -4|\Phi(y,0)|^{2}$ which 
corresponds to the local equilibrium of the wave function and the medium - 
this has an advantage of minimizing transient effects which are unavoidable 
in the system very far from the ground state.  
We also tried other functional forms and values of parameters, and always 
found similar results
as long as the initial functions were spatially well-localized, and the 
wave function had a node. 

We show the results 
of a study using a ring geometry, consisting of $2L = 512$ units 
(larger rings were also 
looked at and the same results were found).  In order to have a fair 
representation of an infinite system we had to make sure that the excitations
of the $w$ field do not travel around the ring, and thus do not interfere 
with the dynamics which would be present in truly infinite system.  This was 
accomplished by adding a localized dissipation source at $y = \pm L$ --  we 
have included a term of the form 
$A\exp[-(y \pm L)^{2}/l^{2}]\partial_{\tau}w$ on the left 
hand 
side of (\ref{newWE}) with adjustable amplitude $A$ and width $l$ ($A = 5$ and
$l = 30$ corresponds to the Figures).  Similarly, 
to eliminate communication between parts of the wave function which are solely
due to the ring geometry, we imposed a very large potential barrier at the 
distances $l$ away from the center of the dissipation source -- this
mimics the vanishing of the wave function at infinity.  These measures 
also turned out to be an effective way to asymptotically slow down and 
confine the soliton fragments in the vicinity of the dissipation region.

In Figure 1 the initial conditions for the electron density and the 
well function were selected to be even ($y_{c} = 0$).  This symmetry is 
preserved by the equations of motion (\ref{newSE}) and (\ref{newWE}) which 
allows us to show the $y < 0$ part of the electron density carpet (left) 
and the $y > 0$ part of the well function (right) on the same plot.  
Ignoring the fine structure, the left and right sides of Figure 1 are 
almost perfect mirror images of each other. Most of the fine structure 
consists of
ripples of the $w$ field which are sound waves propagating with constant 
velocity $s^{-1/2}$.  It is obvious that very quickly after the breakup two
well-localized objects form which travel away from each other, with the 
electron
density and the well function peaks coinciding.  As time progresses, the 
magnitude of the fragment oscillations decreases as they approach the 
Davydov functional form.  This is because the extra kinetic energy 
emitted by the fragments outward in the form of sound waves is carried away 
to infinity -- in our finite system it is absorbed by the dissipative 
segment of the ring opposite to the origin.  This segment can be
recognized on the $w$ carpet as sound-absorbing ripple-free stripe.   
\begin{figure}[htbp]
\epsfxsize=3.2in
\vspace*{-0.1cm}
\epsfbox{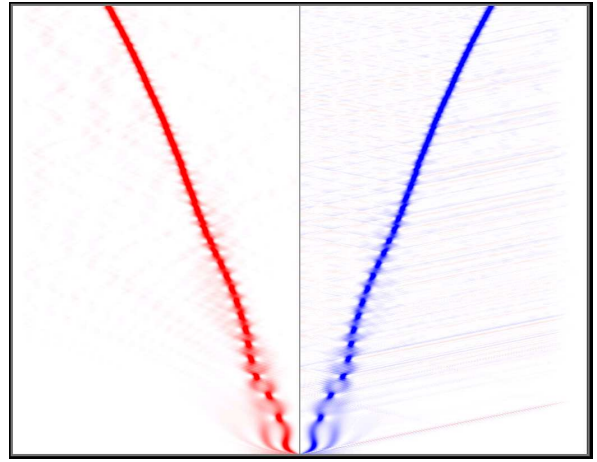}
\vspace*{0.1cm}
\caption{Symmetric splitting of the soliton in half.  For clarity the 
vertical line is drawn to separate the electron density 
(left) and the well function (right) carpets.}
\end{figure} 
It is
clear from Figure 1 that the fragments are slightly accelerating away from 
each other even when they are very far apart.  This is because they 
keep exchanging by sound waves, as can be seen from the $w$ part of the carpet,
which produces an effective repulsion between the fragments.  We tested this
idea by putting at the origin a well-localized source of dissipation of the
same kind as we have at the system edges.  The role of this source is to
prevent the fragments from ``talking'' to each other via sound waves while 
allowing them to
communicate via the wave function.   Then we re-ran the program with 
exactly the same parameters, and turned on the dissipation source half-way
through the evolution.  As a result we observed that after the dissipation 
source was turned on, the fragment trajectories became ballistic thus
confirming that the acceleration was lattice-mediated.  As time progresses 
the efficiency of this effective repulsion decreases as sound waves partially 
transmit through the fragments and leave the system.

For asymmetric initial conditions we observed that the zero of the wave 
function is not preserved during the evolution but the phase discontinuity 
at $y = 0$ persists.  The weak asymmetry cases are fairly similar to the 
symmetric ones and not shown.
Figure 2 shows the $w$ carpet for
a strongly asymmetric breakup of the soliton with the asymmetry parameter 
$y_{c} = 4$ (there is initially substantially more probability on the right 
than on the left). The interesting feature of this evolution is that a 
smaller third fragment persists for quite a long time until it coalesces with
one of the bigger fragments, and we end up with two unequal pieces moving
away from each other.  The density carpet (free of the sound wave ripples) 
looks almost identical to Figure 2 and is not shown.
\begin{figure}[htbp]
\epsfxsize=3.2in
\vspace*{-0.1cm}
\epsfbox{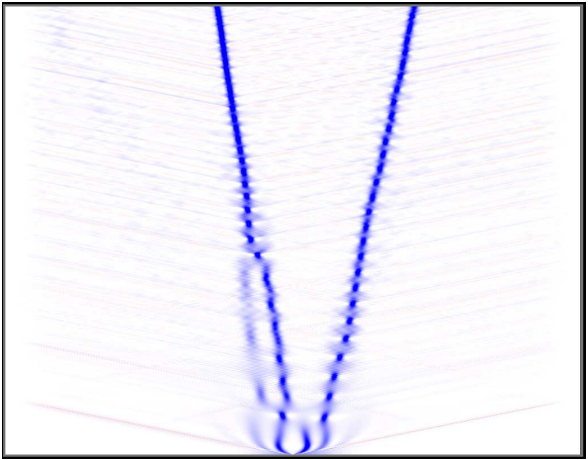}
\vspace*{0.1cm}
\caption{Carpet of the well function for asymmetric splitting of the 
soliton.}
\end{figure} 

We note that in Figures 1 and 2 the velocities of the fragments are 
substantially smaller than the sound velocity which justifies the use of 
linear elasticity theory in describing the medium degrees of freedom.  We also
observed breakups into three and four fragments - this happens for 
smoother initial conditions than those used to
construct Figure 2.  

Our results indicate that the wave function fragments have infinite 
life time.  However a finite large life time cannot be ruled out if the medium
degrees of freedom would be treated quantum-mechanically.    

In interpreting our results it is important to remember  that the 
fissioning of the wave function is a statement about the motion of the 
electron, and not about its parameters such as charge and mass, that 
characterize the electron as a particle \cite{LL2}.  The process which 
leads to the splitting is an example of a {\it measurement} in which the 
quantum system (the electron) interacts with a classical apparatus 
(the medium) \cite{LL2}.  This measurement process changes the state of the 
apparatus and affects the electron by splitting its wave function.  The 
field $w(x,t)$ is the apparatus ``readings'' from which we can draw 
conclusions about the state of the electron, namely the electron probability 
density $|\Phi(x,t)|^{2}$, which can be verified in a subsequent measurement.
If this is a position measurement, a full electron will be found in one of
the fragments, and the wave function will instantly vanish elsewhere. In 
response to that the classical field $w$ in the ``empty'' fragment(s) will 
start relaxing away which will be the experimental sign that before the 
position measurement the electron was in several arbitrarily far separated 
fragments at once.   

We also conducted a similar numerical investigation of the two-dimensional 
version of the problem, and did not find the effect of fissioning which is
so apparent in one dimension.  In two dimensions the electron 
self-localization takes place if the coupling with the medium exceeds some 
critical value, and the polaron is always ``small'' -- its size is of order 
the lattice spacing \cite{Rashba}.  However as long as the observed size of 
the density (or the well) peaks is much larger than the lattice spacing, the 
continuum theory is still valid, and the subsequent conclusions refer to 
this macroscopic regime.  We found that deep 
in the self-localizing regime the two initial density peaks, symmetric with 
respect to a line where the wave function vanishes, do not separate -- they 
just become sharper and remain independent.  The same was also 
true for an asymmetry in the initial conditions. These observations can be 
understood by noticing that above one dimension the amplitude of the sound 
waves is a decreasing function of the distance from the source - the
lattice-mediated interaction is a much weaker effect than in one dimension.
We also observed that when the coupling with the medium is not too large,  
slightly asymmetric peaks have a very different fate -- the higher one 
becomes self-localized, while the lower one diffuses away.  This is 
reminiscent of the scenario put forward by Elser \cite{Elser} in the context 
of electron bubbles in helium. 

This work was supported by the Thomas F. Jeffress and Kate Miller Jeffress 
Memorial Trust.


\begin{references}
\bibitem{note}  The typical time for the electron to change its wave function
is much shorter than the time it takes for heavy particles of the medium to 
adjust to the new wave function.  That is why we can speak of excited states.
\end{references}
\end{document}